\documentclass{IEEEtran4PSCC}

\ifCLASSINFOpdf
   \usepackage[pdftex]{graphicx}
\else
   \usepackage[dvips]{graphicx}
\fi
\usepackage[cmex10]{amsmath}

\makeatletter
\let\old@ps@headings\ps@headings
\let\old@ps@IEEEtitlepagestyle\ps@IEEEtitlepagestyle
\def\psccfooter#1{%
    \def\ps@headings{%
        \old@ps@headings%
        \def\@oddfoot{\strut\hfill#1\hfill\strut}%
        \def\@evenfoot{\strut\hfill#1\hfill\strut}%
    }%
    \def\ps@IEEEtitlepagestyle{%
        \old@ps@IEEEtitlepagestyle%
        \def\@oddfoot{\strut\hfill#1\hfill\strut}%
        \def\@evenfoot{\strut\hfill#1\hfill\strut}%
    }%
    \ps@headings%
}
\makeatother

\psccfooter{%
        \parbox{\textwidth}{\hrulefill \\ \small{Preprint submitted to 22nd Power Systems Computation Conference} \hfill \begin{minipage}{0.2\textwidth}\centering \vspace*{4pt} \includegraphics[scale=0.06]{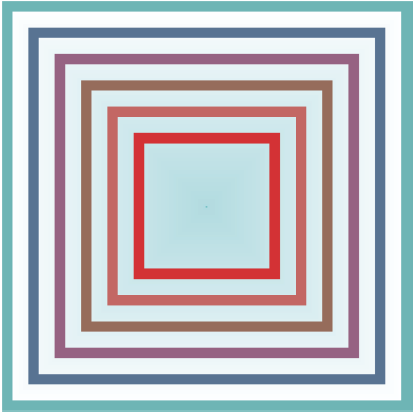}\\\small{PSCC 2022} \end{minipage} \hfill \small{Porto, Portugal --- June 27 -- July 1, 2022}}%
}

\begin{document}

%\title{A numerical tool for non-structure controller aggregation and its application to interconnected Paraguayan-Argentinean power system reduction}

\title{An approach for the aggregation of power system controllers with different topologies}

\author{
\IEEEauthorblockN{Jonas Pesente\\Paulo Galassi}
\IEEEauthorblockA{Electrical Studies Division \\
Itaipu Binacional\\
Foz do Iguaçu, Brazil\\
%\{email author n.1, email author n.2\}@domain (if desired)
}
\and
\IEEEauthorblockN{Leonardo Rodrigues}
\IEEEauthorblockA{Department of Electrical Eng. \\
Unioeste\\
Foz do Iguaçu, Brazil\\
}
\and
\IEEEauthorblockN{Felipe Crestani \\ Guilherme Justino}
\IEEEauthorblockA{AS.DT \\
Itaipu Technological Park\\
Foz do Iguaçu, Brazil\\
}
\and
\IEEEauthorblockN{Rodrigo Ramos}
\IEEEauthorblockA{Department of Electrical Engineering \\
Engineering School of Sao Carlos \\
University of Sao Paulo\\
Sao Carlos, Brazil\\
}
}

\maketitle

\begin{abstract}
This paper proposes an approach to aggregate non-structured power system controllers preserving the dynamical characteristics of the original devices. The method is based on linear operations that use the frequency response of the elements, resulting in an accurate input-output description of the equivalent controller when compared to the original ones. The developed method was applied to a model of the future interconnected Paraguayan-Argentinean power system to produce a dynamic equivalent used in a real-time simulator to test the special protection scheme needed for the safe operation of the this future system. Transient and small-signal stability studies presented matching simulation results in the time domain with significantly reduced computational burden and processing time.
\end{abstract}


\thanksto{\noindent Submitted to the 22nd Power Systems Computation Conference (PSCC 2022).}

%\vspace{-1cm}

\section{Introduction}
Several techniques to obtain reduced models with equivalent representations of the dynamics of large electric power systems have been developed by the scientific community to decrease the computational cost of simulation and allow its implementation in real-time simulators for Intelligent Electronic Device (IED) testing and controller certification through closed-loop simulations.

These techniques can be classified into three main approaches \cite{annakkage2011dynamic}: (a) aggregation of dynamical elements with similar dynamical responses by identifying coherent generators (\cite{ourari2006dynamic,anaparthi2005coherency}), (b) system reduction based on linearized models (\cite{paternina2016dynamic,wang2014dynamic}), and (c) system identification using on-line measurements (\cite{ju2004dynamic1,ju2004dynamic2}).

The techniques based on computing an equivalent element of each group of coherent generators can preserve their physical structure. However, when it comes to their controllers, the available bibliography only presents methods to compute an equivalent if the controllers of interest have sets of input and output signals that are identical (i.e., have the same physical quantities) (\cite{zin2003time,ourari2006dynamic}). To the extent of the knowledge of the authors, there is a lack of methods to aggregate controllers with different topologies, for instance, Power System Stabilizers (PSS) based on rotor speed ($\Delta\omega$), active power ($\Delta P_E$), or a combination of these two measurements.

To fill this gap, the main contribution of this paper is a general formulation for the aggregation of class-alike controllers with different input or output signals.

The technique was applied to compute a dynamic equivalent model for the future Paraguayan-Argentinean (PY-AR) interconnected power system, presenting reliable results that allowed its use in hardware-in-the-loop (HIL) tests of the special protection scheme designed for its safe operation. The model was then implemented on a real-time digital simulator, and its logic functions were tested and fine-tuned. The paper details the steps for its computation and the results obtained.

This paper is structured as follows: Section II presents a method with all steps to produce an equivalent model for the PY-AR system; Section III describes the main proposal of this paper, intended to overcome a limitation of the method presented in Section II; Section III shows the results of computational experiments performed with the application of the proposed method; and Section IV finishes the paper with concluding remarks.   

%---------------------------------------------------------------------------------------------
\section{Development of a Dynamic Equivalent Model for the PY-AR System}

This section describes a set of steps necessary to build a dynamic equivalent model for the future PY-AR power system based on \cite{zin2003time}. The idea is to use this equivalent model as a basis for comparison with the results given by the method proposed in this paper, so the advantages of the latter can be highlighted. 

\begin{figure*}[h]
% \hspace*{-0.6cm}
\centering
    \includegraphics[scale=0.63]{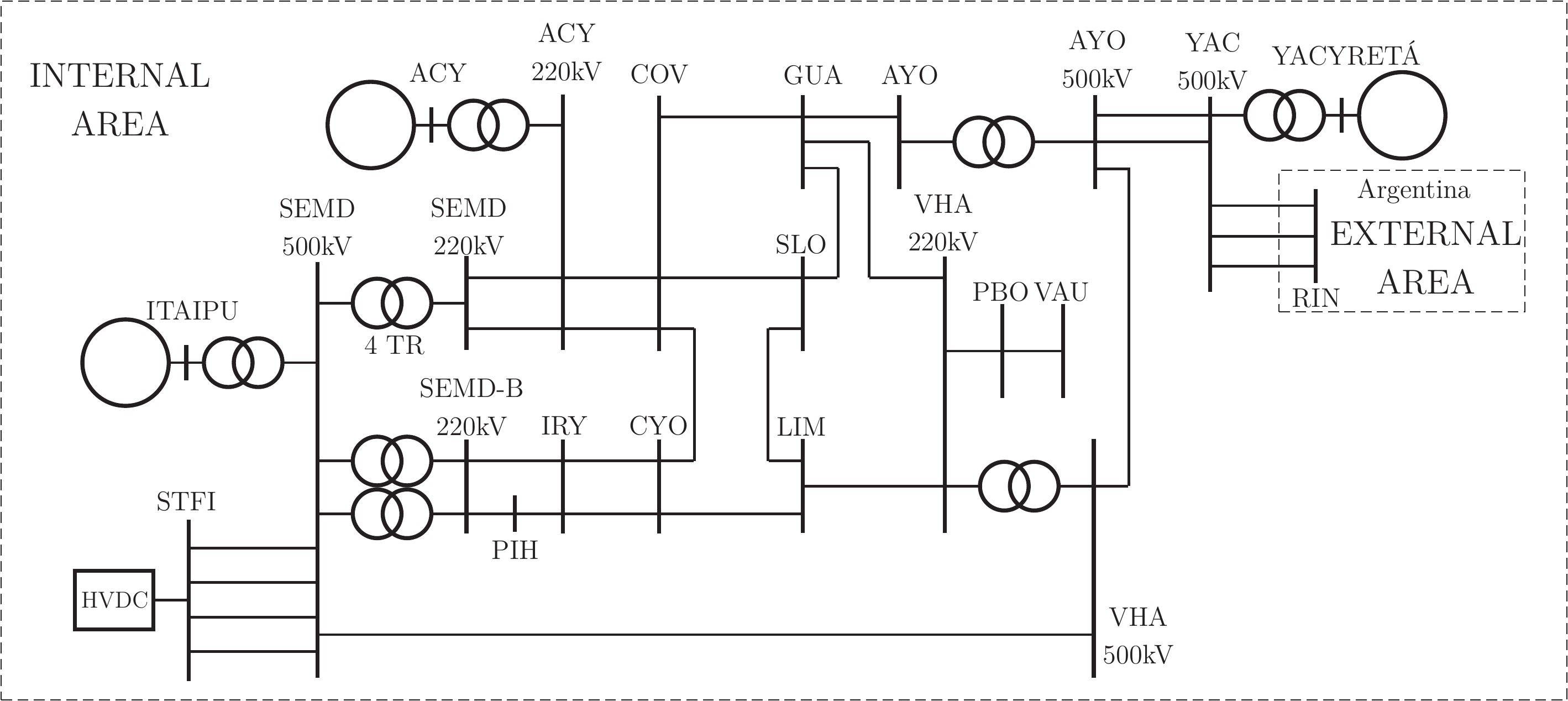}
     \caption{One-line diagram of the future PY-AR system with the definitions of the internal and external areas. }
     \label{fig:fig01}
\end{figure*}

The one-line diagram of the future PY-AR system is shown in  Fig. \ref{fig:fig01}. The adjective ``future'' is used to denote the fact that the PY and AR system are not yet interconnected, although this interconnection is about to happen. 

% Figura realocada

The following steps summarize the procedure applied to build the equivalent that will be used as a basis of comparison in this paper:

\begin{enumerate}
    \item Definition of the study area, or \textit{internal area}, which is the portion of the network that is fully retained. It is also in this area where contingencies (events) will be carried out and where the effects of disturbances are studied;
    \item Identification and grouping of coherent generators of the \textit{external area};
    \item Reduction of the static model of the transmission network in two stages: (i) calculation of the Radial Equivalent Independent (REI) meshes; and (ii) elimination of the load buses;
    \item Dynamic aggregation of coherent generating units \cite{podmore1978identification}, which can be summarized as obtaining the parameter values for the equivalent units from each coherent group of generators and their respective controls.
\end{enumerate}

The application of these steps to obtain an equivalent for the future PY-AR power system is detailed in the following subsections. The final purpose of the development of dynamic equivalents of this system was to perform Hardware-In-the-Loop (HIL) tests in the Special Protection Scheme (SPS) developed to ensure safety to the operation of Paraguayan and Argentinean power systems in an interconnected condition \cite{pesente2018}, \cite{pesente2019}, \cite{pesente2021}. 

\subsection{Definition of the study region}

The study area defines the region of the system that is kept in detail. In this step, the buses connecting the system to be studied and the system to be simplified are defined as the boundaries between the study area and the external area. The essential equipment models kept for the PY-AR equivalent are presented in Fig. \ref{fig:fig01}, as well the boundaries defined for the model (the interface YAC and RIN buses).

% Figura estava aqui

\subsection{Identification of coherent generators}

Groups of coherent generating units can be identified by analyzing their responses to disturbances, for example. Once identified, an external system equivalent can be built by replacing the coherent groups with an equivalent machine model. 

\begin{figure}[ht]
% \hspace*{-0.6cm}
\centering
    \includegraphics[scale=0.82]{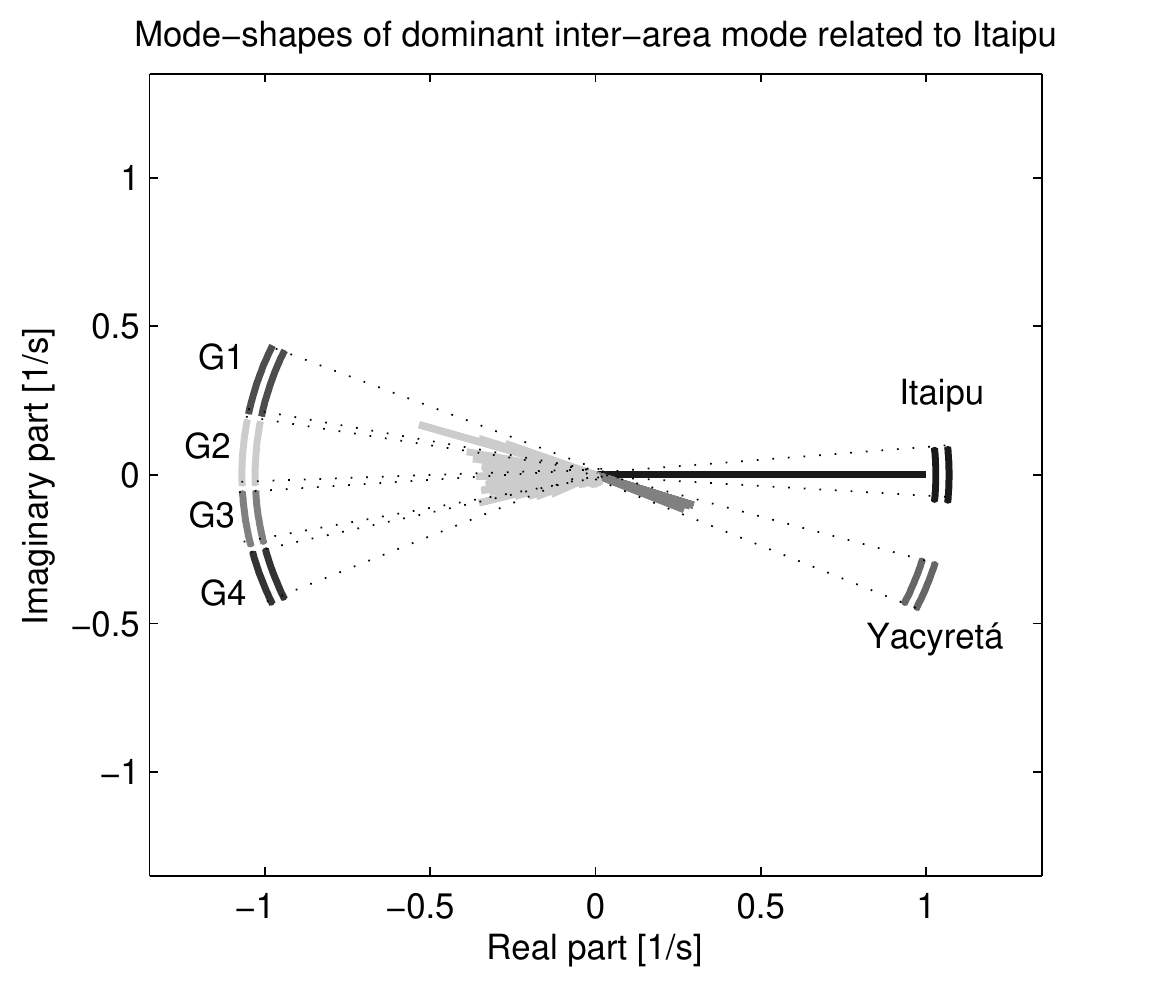}
     \caption{Definition of coherent groups for a specific oscillation mode.}
     \label{fig:Fig02md}
\end{figure}

This work uses a modal approach based on a linearized model of the future PY-AR system, where its eigenstructure is used to define the coherent generator groups, as follows:

\begin{enumerate}
    \item Calculate the system eigenvalues and select the ones corresponding to the critical modes of the system;
    \item Calculate the mode shapes (eigenvectors) related to the critical modes and compare the entries of these vectors related to the rotor speeds of the generators ($\Delta \omega$), which are displayed in Fig. \ref{fig:Fig02md};
    \item Group the generators by similarity in the angles of the respective entries compared in the previous step, which leads to the groups denoted in Fig. \ref{fig:Fig02md} as G1, G2, G3, G4, Yaciretá, and Itaipu  \cite{han1995discovery}.
\end{enumerate}

\subsection{Reduction of the static network model}
\begin{figure}[ht]
% \hspace*{-0.6cm}
\centering
    \includegraphics[scale=0.53]{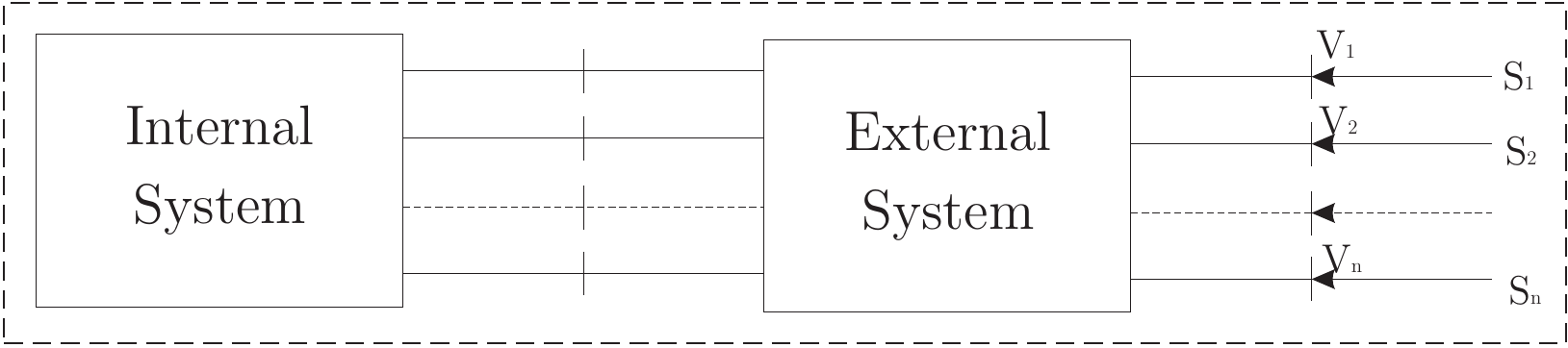}
     \caption{Separate representation of the generators in a coherent group before REI equivalencing.}
     \label{fig:fig02-1}
\end{figure}

\begin{figure}[ht]
% \hspace*{-0.6cm}
\centering
    \includegraphics[scale=0.63]{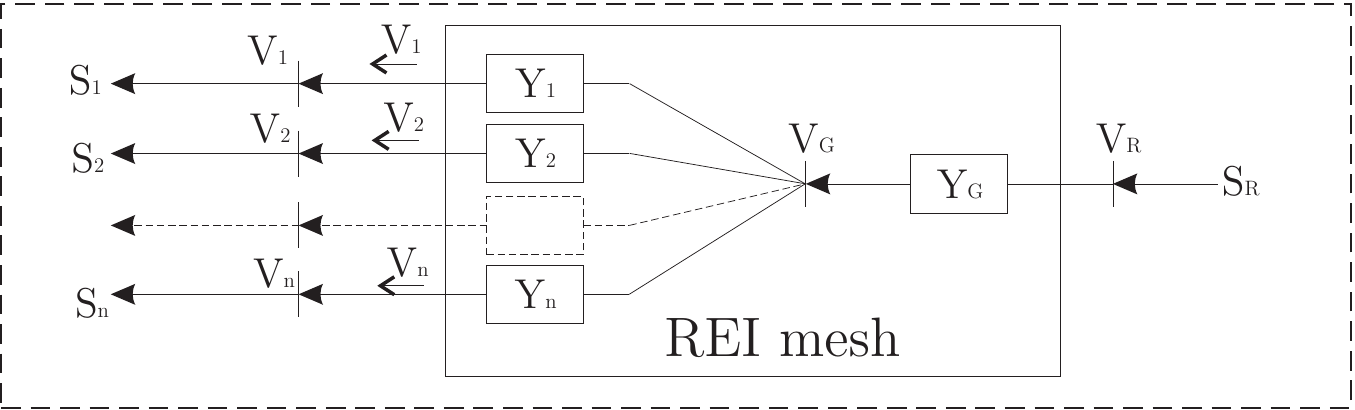}
     \caption{Equivalent representation of the generators in a coherent group after REI equivalencing.}
     \label{fig:fig02-2}
\end{figure}

First, the generators of each coherent group are allocated to a common terminal busbar. Then, using the REI formulation, the power injections of the coherent generators in each group are transferred to a radial equivalent mesh, which is independent of the remaining of the network. The process is illustrated by the transition from Fig. \ref{fig:fig02-1} to Fig. \ref{fig:fig02-2}.

This method preserves key features of the representation of generators, such as the operation as a controlled voltage source, for example. Furthermore, its resulting network model is suitable to the RMS dynamic formulations typically employed in stability studies of large-scale power systems.

\subsection{Dynamic Model Aggregation}

\vspace{0.4cm}

\subsubsection{Synchronous generators}\hfill\\

The dynamic aggregation of coherent generating units grouped in the same bus consists in determining the parameters of a single equivalent generating unit. For this purpose, the aggregation technique assumes that the coherent units operate in parallel.

This work applied the formulation presented in \cite{zin2003time}, in which the determination of the equivalent reactance uses relations among currents and voltages of the individual coherent units. 

The angle difference among these individual coherent units with respect to the common reference (assumed to be the angle of the voltage at the bus where the equivalent is connected) is also considered. Finally, the equivalent inertia constant is calculated by the sum of the individual inertia constants of all coherent units.\\

\subsubsection{Control systems}\hfill\\

The computation of equivalents for control system of coherent synchronous generators is a difficult task. One of the concepts on which this task can be based is that the equivalent controller must be such that it produces on the equivalent machine the same effect that the original controllers would produce on the coherent generators.

From multiple perspectives, the concept described in the previous paragraph can be considered as ill-defined. This is one of the motivations for the proposal of this paper, which will be described in the next section. 

However, to build the basis of comparison, it is possible to better define the mentioned concept by assuming that all controllers being aggregated have the same input and output quantities. Then, it is possible to compute an Aggregate Transfer Function (ATF) with the same Frequency Response (FR) of the sum of the FRs of each individual controller, weighted by the rated power of its respective generator. This work uses the least squares method to compute the ATF for this task.

\begin{figure}[ht]
% \hspace*{-0.6cm}
\centering
    \includegraphics[scale=0.37]{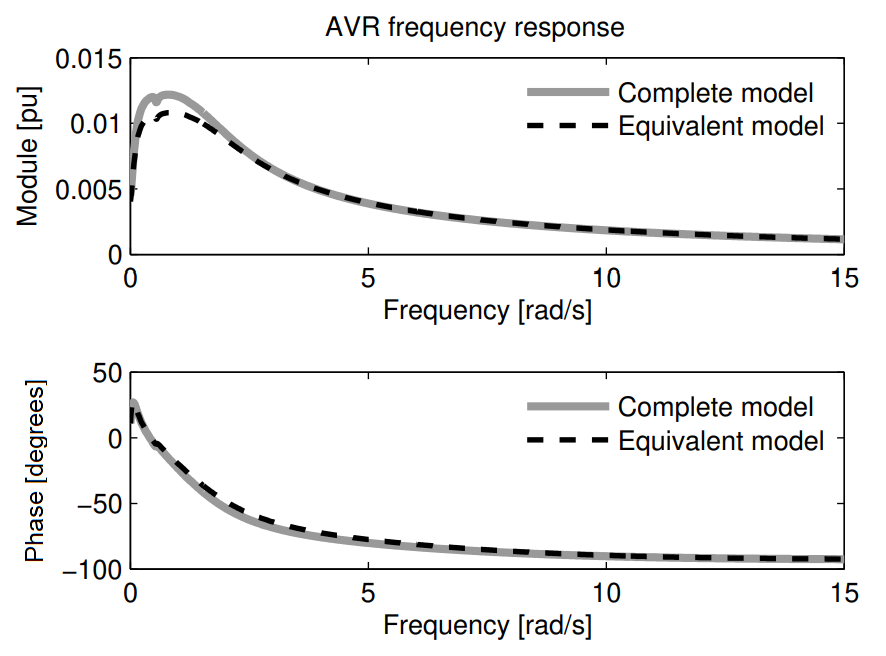}
    \caption{Comparison of the aggregated frequency response of AVRs and the frequency response of the TF equivalent.}
     \label{fig:fig04-1}
\end{figure}

Fig. \ref{fig:fig04-1} illustrates the FR of an ATF computed from two Automatic Voltage Regulators (AVR) of two parallel generators compared to the sum of their weighted FRs.

% PHASE INSTEAD OF ANGLE AND MODULE IN THE BOTTOM FIGURE!!!

\begin{figure}[ht]
% \hspace*{-0.6cm}
\centering
    \includegraphics[scale=0.59]{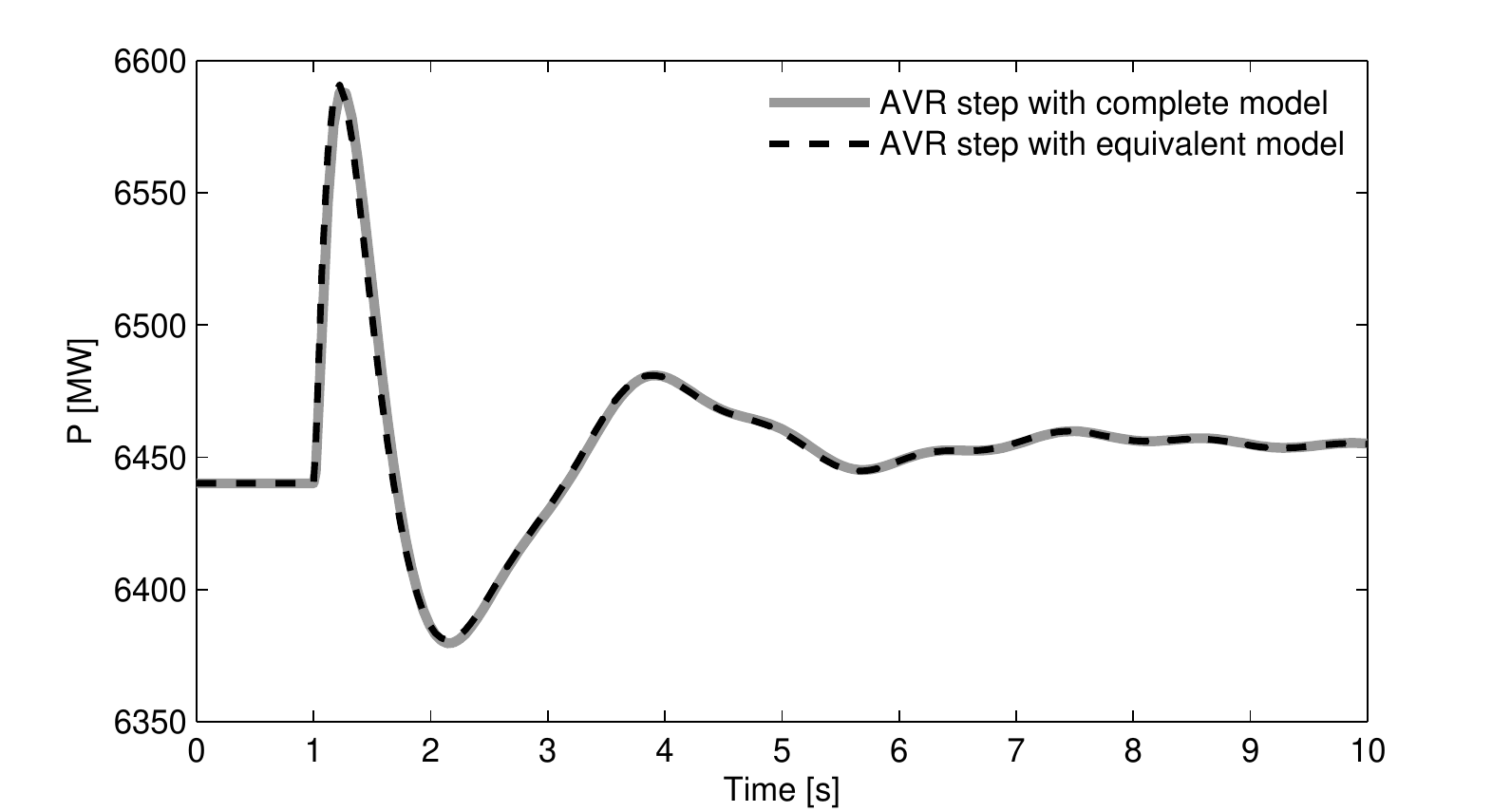}
     \caption{Comparison of the time response for the original and equivalent controllers.}
     \label{fig:fig04-2}
\end{figure}

Similarly, Fig. \ref{fig:fig04-2} illustrates the corresponding time responses of the sum of the active power of the two generators when compared to the active power of the equivalent generator equipped with the equivalent AVR.

Figs. \ref{fig:fig04-1} and \ref{fig:fig04-2} show the excellent performance of this technique on the aggregation of control systems when the premise that the individual controllers have the same input and output quantities holds. For the cases where this premise does not hold, the next section presents the main proposal of this paper.

%--------------------------------------------------------------------------------------------- JJJJ

\section{Proposal for the aggregation of controllers with different topologies}\label{SecAgg}

The purpose of this section (which is also the main contribution of this work) is explained in the following paragraphs. 

Although the method for controller aggregation presented in the previous section exhibited good results, it would not work if the controllers to be aggregated did not have input and output variables of the same nature. To overcome this drawback, this paper proposes the method described in this section, which extends the same idea of the previous method to the case where the inputs and outputs of the controllers to be aggregated have different physical natures. 

\begin{figure}[ht]
% \hspace*{-0.6cm}
\centering
    \includegraphics[scale=1.1]{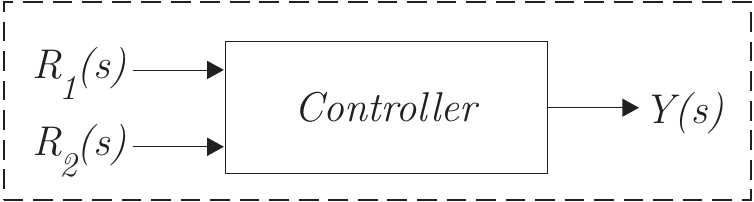}
     \caption{Block diagram of a general controller with two inputs and one output}
     \label{fig:fig0X}
\end{figure}

For simplicity, the formulation of the proposed method is based on a general controller with two inputs and one output presented in Fig. \ref{fig:fig0X}.

Assuming that the equivalent controller structure for the group of controllers is represented by the relation $H_1(s)=Y(s)/R_1(s)$ and that there is a need to include a particular controller with a topology defined by the transfer function $H_2(s)=Y(s)/R_2(s)$.

This particular controller $H_2(s)$ can be included in the equivalent controller structure using the following set of steps:

\vspace{0.3cm}

\begin{itemize}
    \item Compute the frequency response (for the range of interest) of the particular controller to be aggregated as  $H_2(s)=Y(s)/R_2(s)$;
    \vspace{0.2cm}
    \item Define the transfer function between the inputs of the particular controller and the current equivalent controller as $H_{21}(s)=R_2(s)/R_1(s)$;
    \vspace{0.2cm}
    \item Compute the frequency response of $H_{21}(s)$ for the range of interest;
    \vspace{0.2cm}
    \item Given that the relation $Y(s)=H_1(s)R_1(s)=H_2(s)R_2(s)$ implies that $H_1(s)=H_2(s)(R_2(s)/R_1(s))$. the frequency response of the final aggregated controller can be calculated by $H_1(s)=H^{-1}_{21}(s)H_2(s)$ and its computation requires only the knowledge of $H_{21}(s)$ and $H_2(s)$, which were computed in the previous steps. Note that $H_2(s)$ must have been previously weighted by the rated power of its respective generator to result in the correct contribution to its group.
\end{itemize}
\vspace{0.2cm}

Also, the proposed method can be easily adapted to a general controller similar to the one shown in Fig. \ref{fig:fig0X}, with a single input and two outputs. Furthermore, the proposed method and its mentioned adaptation can be applied sequentially, therefore being applicable to the aggregation of controllers with multiple inputs and outputs of different physical nature.   It is important to remark that this sequential application assumes that all of the corresponding transfer functions describe single-input single-output relations.

An example of the application of this proposed method is presented in the next section.   

\section{Computational Experiments}

The computational experiments of this work used the proposed approach to develop an equivalent dynamic model to enable test of devices that will compose the interconnection of the PY and AR power systems. The PY-AR interconnection is expected to happen in 2021.

Studies performed to evaluate this interconnection have shown a potential occurrence of an unstable inter-area oscillation in the range of $0.31-037$Hz in this PY-AR interconnected model with the participation of the Itaipu power plant \cite{estudos1}.

Therefore, the equivalent developed has the priority to retain this oscillation mode in order to help finding the best strategies to avoid problems associated with this dynamic behavior.

The investigation of this particular oscillation mode revealed a strong coherent behavior of the generators of the system for disturbances inside the Paraguayan power system, as presented in Fig. \ref{fig:Fig02md}. 

In Fig. \ref{fig:Fig02md} it is possible to verify four groups of generators in the AR system that oscillate against the Itaipu and Yacyretá power plants (which are retained in the internal system). 

\subsection{An equivalent controller example}

One of the challenges presented in developing the PY-AR dynamic equivalent was related the PSS structure of aggregate generators in the equivalent models. This occurs because there are different types of PSS in  generators within each of the 4 groups. The most common structures found in the complete model were based on inputs $\Delta \omega$, $\Delta P_E$, or a combination of $\Delta \omega$ and $\Delta P_E$.

To aggregate these PSS in one equivalent controller, this work adopted a PSS structure based on a $\Delta \omega$ input. After that, all the PSS based on $\Delta P_E$ or $\Delta \omega$ and $\Delta P_E$ as inputs were transformed into $\Delta \omega$ input-type PSS.

Based on the procedure proposed in Section \ref{SecAgg}, all $\Delta P_E$ inputs could be included in the equivalent PSS structure by computing the frequency response of  $\Delta \omega/\Delta P_E$. Then it was multiplied by the frequency response of $\Delta V_{PSS-PE}/\Delta P_E$, resulting in a path between $\Delta V_{PSS-PE}/\Delta \omega$ that includes the contribution of the $\Delta V_{PSS-PE}/\Delta P_E$ PSS added to the aggregated model.

\begin{figure}[h]
% \hspace*{-0.6cm}
\centering
    \includegraphics[scale=0.7]{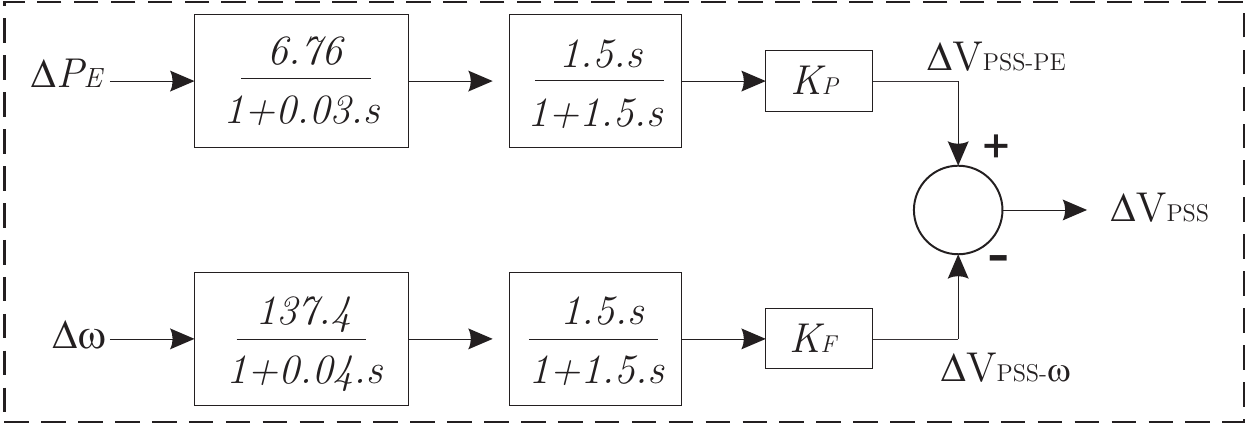}
     \caption{Original PSS structure of the Itaipu generators}
     \label{fig:fig07}
\end{figure}

\begin{figure}[h]
% \hspace*{-0.6cm}
\centering
    \includegraphics[scale=0.55]{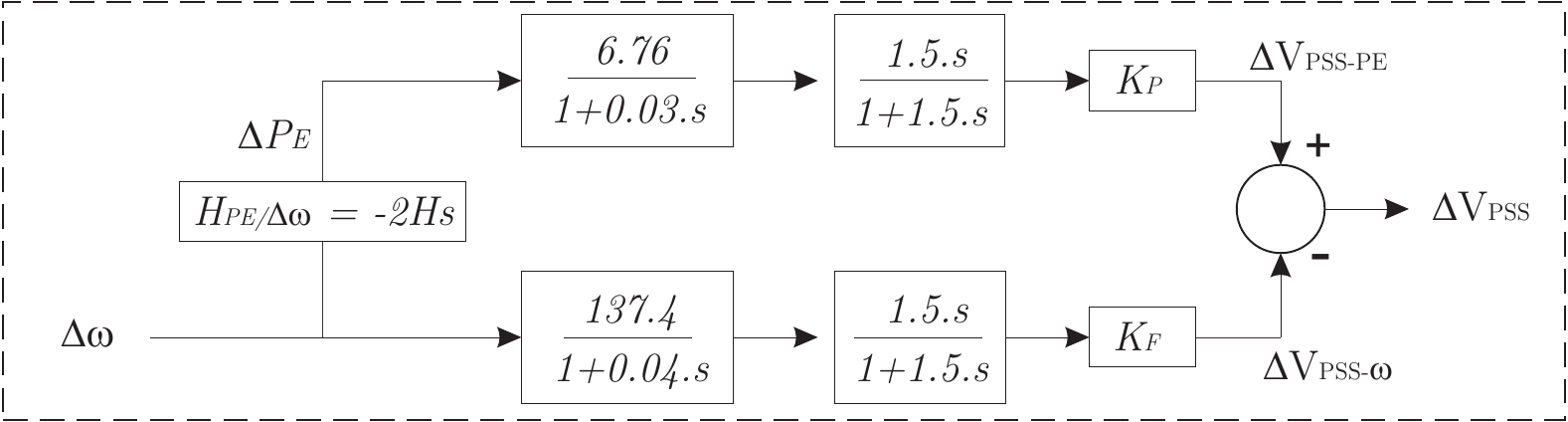}
     \caption{Equivalent PSS structure of the Itaipu generators}
     \label{fig:fig08}
\end{figure}

In the following, an illustration of the transformation of a two-input into a one-input stabilizer with the application of the procedure proposed in Section \ref{SecAgg} is presented for the PSS of Itaipu. Fig. \ref{fig:fig07} presents the original Itaipu PSS dynamic blocks and Fig. \ref{fig:fig08} presents the aggregated PSS based only on a $\Delta \omega$ input.

Although the Itaipu generators are retained in the internal area (and, therefore, there is no need to calculate an equivalent for them in this application), this specific case is quite interesting because the resulting equivalent PSS can be described by the addition of a path with a very intuitive physical meaning. From a small disturbance viewpoint and considering a fixed mechanical power, the frequency response of $\Delta \omega/\Delta P_E$ is given by $-2Hs$, which exactly matches the result obtained with the application of the proposed method. 

Therefore, the analytical expression shown in Eq. (\ref{eq:eq01}) describes the equivalent Itaipu PSS only in terms of its $\Delta \omega$ input. The inertia constant of the Itaipu generators is equal to 5.07MVA/MWs.

\begin{equation}
    \frac{V_{PSS}}{-\Delta \omega}=\left(\frac{1.5s}{1.5s+1}\right)\left(\frac{6.869s^2+205.9s}{0.0012s^2+0.07s+1}\right)
    \label{eq:eq01}
\end{equation}

\begin{figure}[ht]
% \hspace*{-0.6cm}
\centering
    \includegraphics[scale=0.6]{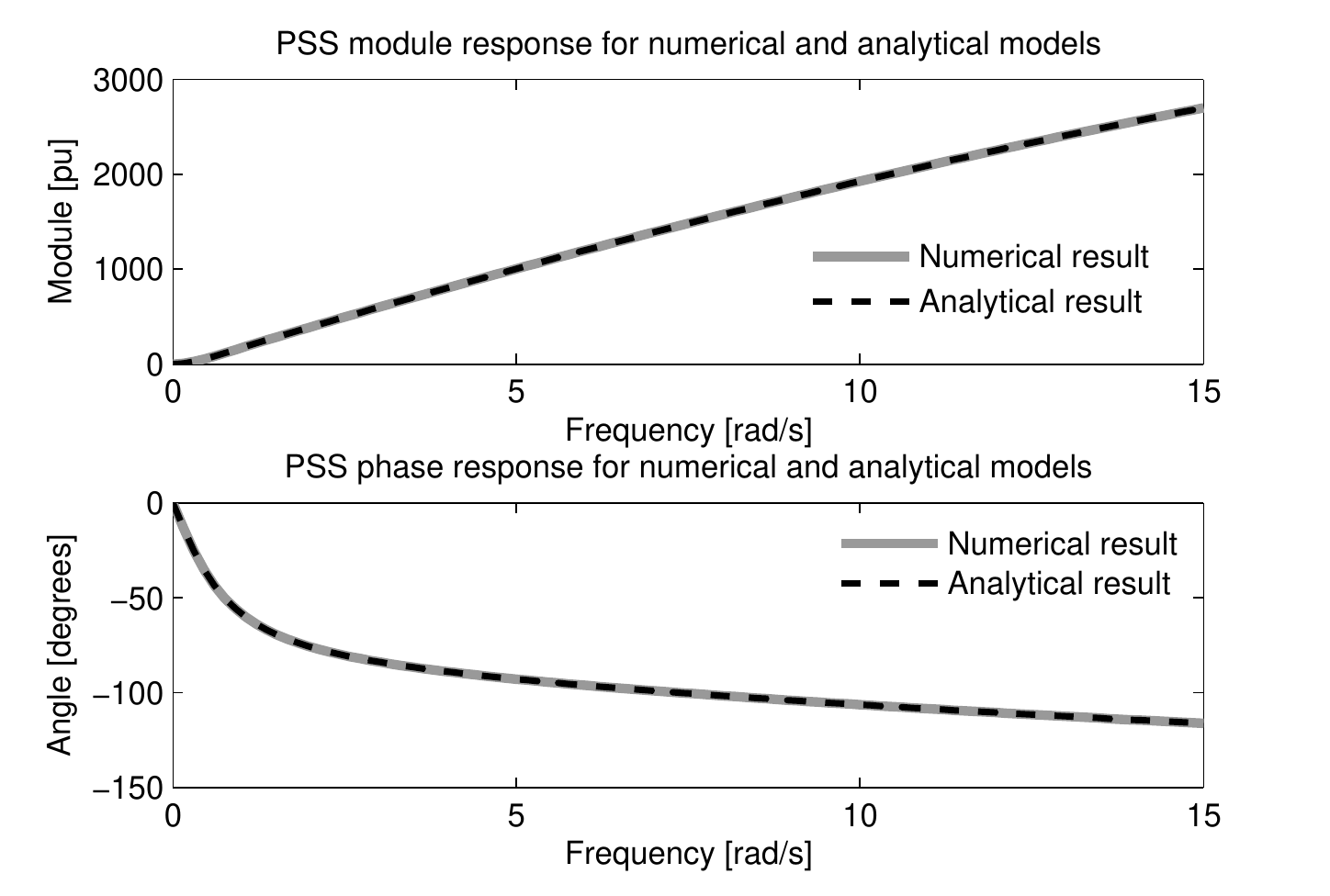}
     \caption{PSS frequency response for analytical and numerical methods}
     \label{fig:fig09}
\end{figure}

Fig. \ref{fig:fig09} presents the comparison of the frequency responses resulting from the analytical expression and the numerical computation with the proposed method, showing the correspondence between both approaches and validating the proposal.

It is important to remark the this double-inputs transformation into a single-input equivalent controller can also be (and indeed was) performed for governors that presented $\Delta P_E$ and $\Delta \omega$ inputs, giving extra results to validate the proposed method.  

\subsection{The PY-AR equivalent: performance under large disturbances}

Table \ref{tab:tab01_stat} presents some important features of the complete and the equivalent models.

\begin{table}[ht]
\caption{Features of the complete and equivalent PY-AR models}
\begin{tabular}{lcc}
\hline
\textbf{Number of}            & \textbf{Complete model} & \textbf{Equivalent model} \\ \hline
Buses                         & 3849                    & 600                       \\ \hline
Lines                         & 2682                    & 411                       \\ \hline
Transformers                  & 2507                    & 467                       \\ \hline
Generators                    & 292                     & 7                         \\ \hline
Controllers                   & 307                     & 26                        \\ \hline
Individual loads              & 225                     & 0                         \\ \hline
Shunt bus-connected           & 124                     & 88                        \\ \hline
Shunt line-connected          & 110                     & 41                        \\ \hline
Static compensator            & 5                       & 5                         \\ \hline
Shunt bank                    & 331                     & 11                        \\ \hline
Special Protection Schemes    & 24                      & 0                         \\ \hline
Protection models and relays  & 118                     & 0                         \\ \hline
\end{tabular}
\label{tab:tab01_stat}
\end{table}

%JJJJJJJ
This work considered four steady-state conditions to represent the system operating limits for studies under large disturbances, as presented in Table \ref{tab:tab02_SS}.

\vspace{-0.3cm}

\begin{center}
\begin{table}[ht]
\caption{Steady-state cases}
\begin{tabular}{lccc}
\hline
\textbf{Case} & \textbf{PY-Load {[}MW{]}} & \textbf{Itaipu-PY {[}MW{]}} & \textbf{Yacyreta-PY {[}MW{]}} \\ \hline
1             & 3.850                     & 2.100 (55\%)                & 1.350 (35\%)                  \\ \hline
2             & 3.850                     & 2.100 (55\%)                & 500 (13\%)                    \\ \hline
3             & 1.490                     & 850 (57\%)                  & 500 (34\%)                    \\ \hline
4             & 1.490                     & 1.050 (70\%)                & 280 (19\%)                    \\ \hline
\end{tabular}
\end{table}    
\label{tab:tab02_SS}
\end{center}

\vspace{-0.5cm}

The average time taken for the simulation of a 50s time window was 44min and 52s using the complete model and 1 min and 2 s using the equivalent one, which shows the importance of the equivalent representation to enable batch studies of the system. 

With respect to the precision of the results, two cases illustrate the equivalent model response when compared to complete model response, both for a 100ms short-circuit in the SEMD 500 kV substation (see Fig. \ref{fig:fig01}).

\begin{figure}[ht]
% \hspace*{-0.6cm}
\centering
    \includegraphics[scale=0.65]{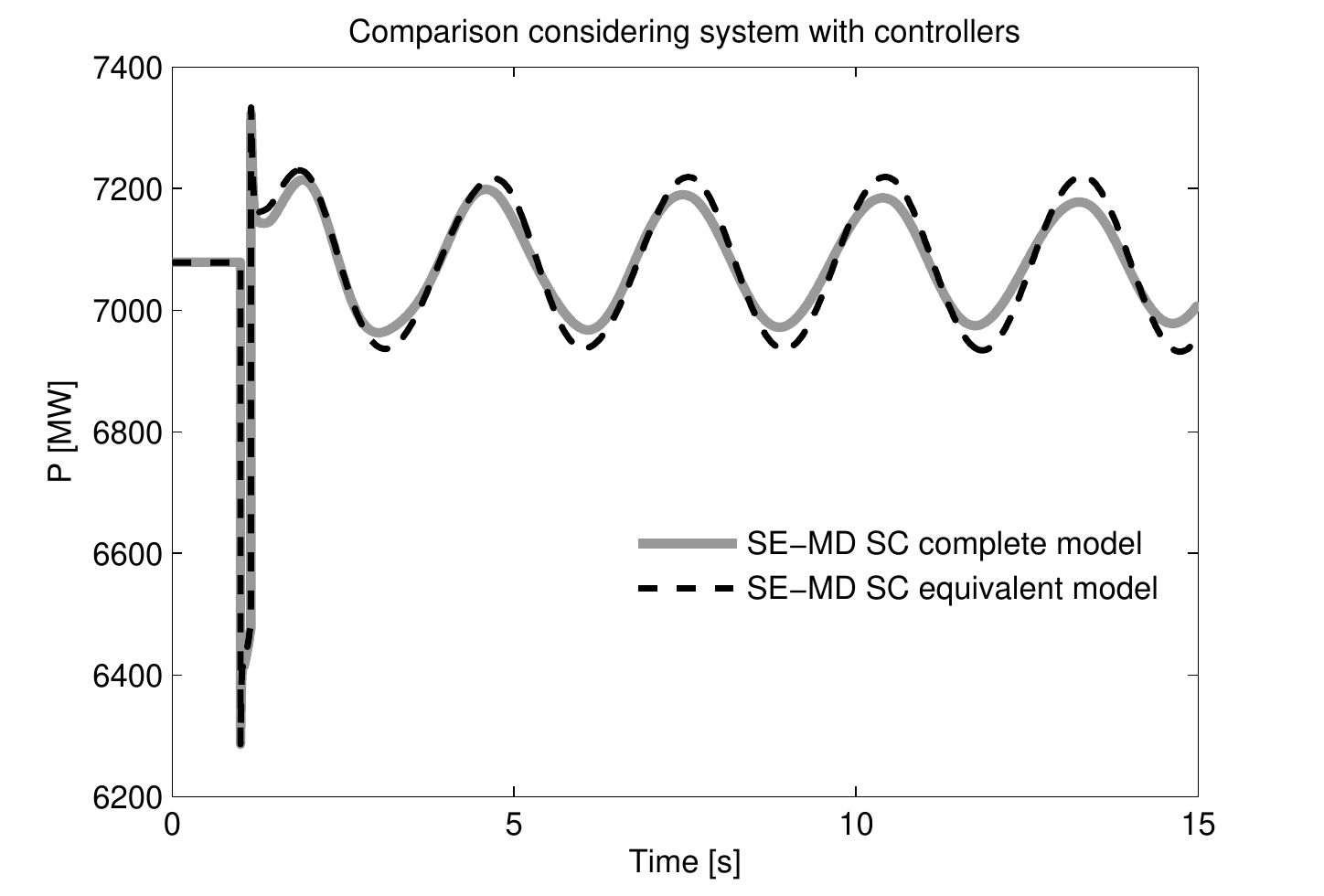}
     \caption{Active power of Itaipu for a 100ms short-circuit in SEMD substation, without controllers}
     \label{fig:fig10}
\end{figure}

The first one presented in Fig. \ref{fig:fig10} considers all generators without controllers to emphasize the quality of the network reduction and dynamic aggregation of generators. The results in Fig. \ref{fig:fig10} show a good compatibility of the responses between the responses of the equivalent and the complete models, although there is still some level of imprecision.

\begin{figure}[ht]
% \hspace*{-0.6cm}
\centering
    \includegraphics[scale=0.65]{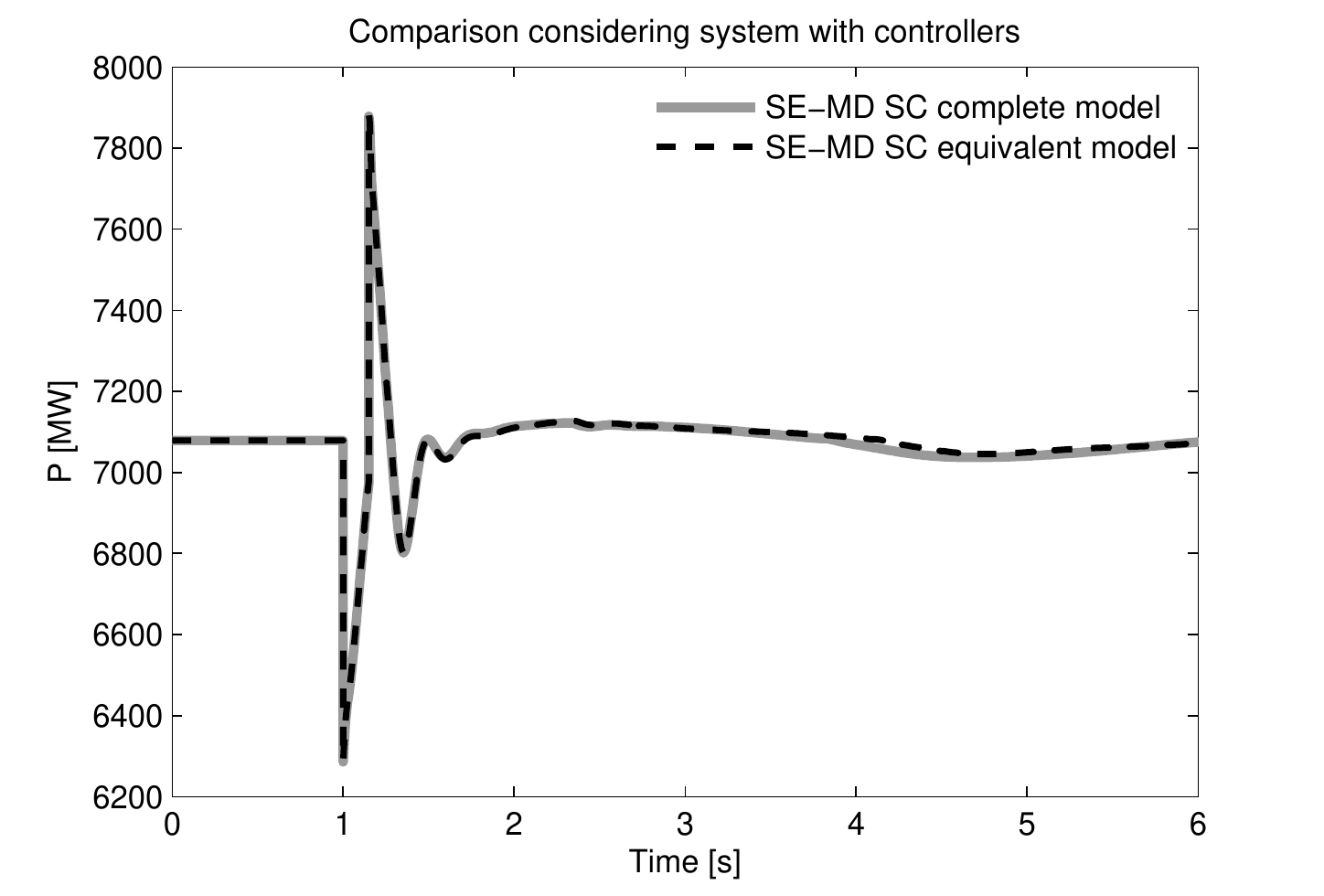}
     \caption{Active power of Itaipu for a 100ms short-circuit in SEMD substation, with controllers}
     \label{fig:fig11}
\end{figure}

The second one, presented in Fig. \ref{fig:fig11} considers all the controllers of the system, to emphasize the precision of the proposed method for controller aggregation.

Therefore, the results in this section show that the PY-AR equivalent model obtained with the application of the proposed method enable a fast and precise simulation of the complete system response. 

\subsection{The PY-AR equivalent: performance under small disturbances}

The evaluation of the proposed procedure also considered the viewpoint of small perturbations. Table \ref{tab:tab03_stat} presents the dominant inter-area mode for the complete and the equivalent models considering all power system controllers.

\begin{table}[ht]
\centering
\caption{Results of small-signal analysis for the complete and equivalent models}
\begin{tabular}{ccccc}
\hline
\textbf{Case}   & \multicolumn{2}{l}{\textbf{Complete model}} & \multicolumn{2}{l}{\textbf{Equivalent model}} \\ \hline
                & f[Hz]   & $\xi$ [\%]      & f[Hz]    & $\xi$ [\%] \\ \hline
1               & 0.307   & 14.2            & 0.312    & 21.3       \\ \hline
2               & 0.291   & 15.0            & 0.260    & 16.6       \\ \hline
3               & 0.320   & 11.1            & 0.319    & 21.2       \\ \hline
4               & 0.317   & 11.2            & 0.304    & 20.8       \\ \hline
\end{tabular}
\label{tab:tab03_stat}
\end{table}

Table \ref{tab:tab03_stat} shows that the equivalent model was able to retain the dominant mode in all 4 cases, with a maximum numerical difference of 5.1\% for the damping factor for case 3. This shows the adequacy of the proposed procedure for applications involving small-signal stability studies. 

%---------------------------------------------------
\subsection{The PY-AR equivalent for real-time simulation}

For real-time and hardware-in-the-loop simulations, the restrictions under the maximum size of the external system were even stricter than the ones for computer simulation. A 22-bus equivalent, shown in Fig. \ref{fig:fig01}, was developed for this purpose. Table \ref{tab:tab04_stat} presents the main features of this 22-bus equivalent model, which can be compared to the ones given in Table \ref{tab:tab01_stat}.

\begin{table}[ht]
\centering
\caption{Features of the model for real-time simulation}
\begin{tabular}{lc}
\hline
\textbf{Number of}            & \textbf{Equivalent model} \\ \hline
Buses                         & 22                        \\ \hline
Lines                         & 28                        \\ \hline
Transformers                  & 8                         \\ \hline
Generators                    & 3                         \\ \hline
Controllers                   & 9                         \\ \hline
Individual loads              & 0                         \\ \hline
Shunt bus-connected           & 3                         \\ \hline
Shunt line-connected          & 0                         \\ \hline
Static compensator            & 0                         \\ \hline
Shunt bank                    & 3                         \\ \hline
Special Protection Schemes    & 0                         \\ \hline
Protection models and relays  & 0                         \\ \hline
Average simulation time (50s) & 1.64s                     \\ \hline
\end{tabular}
\label{tab:tab04_stat}
\end{table}

\begin{figure}[ht]
% \hspace*{-0.6cm}
\centering
    \includegraphics[scale=0.67]{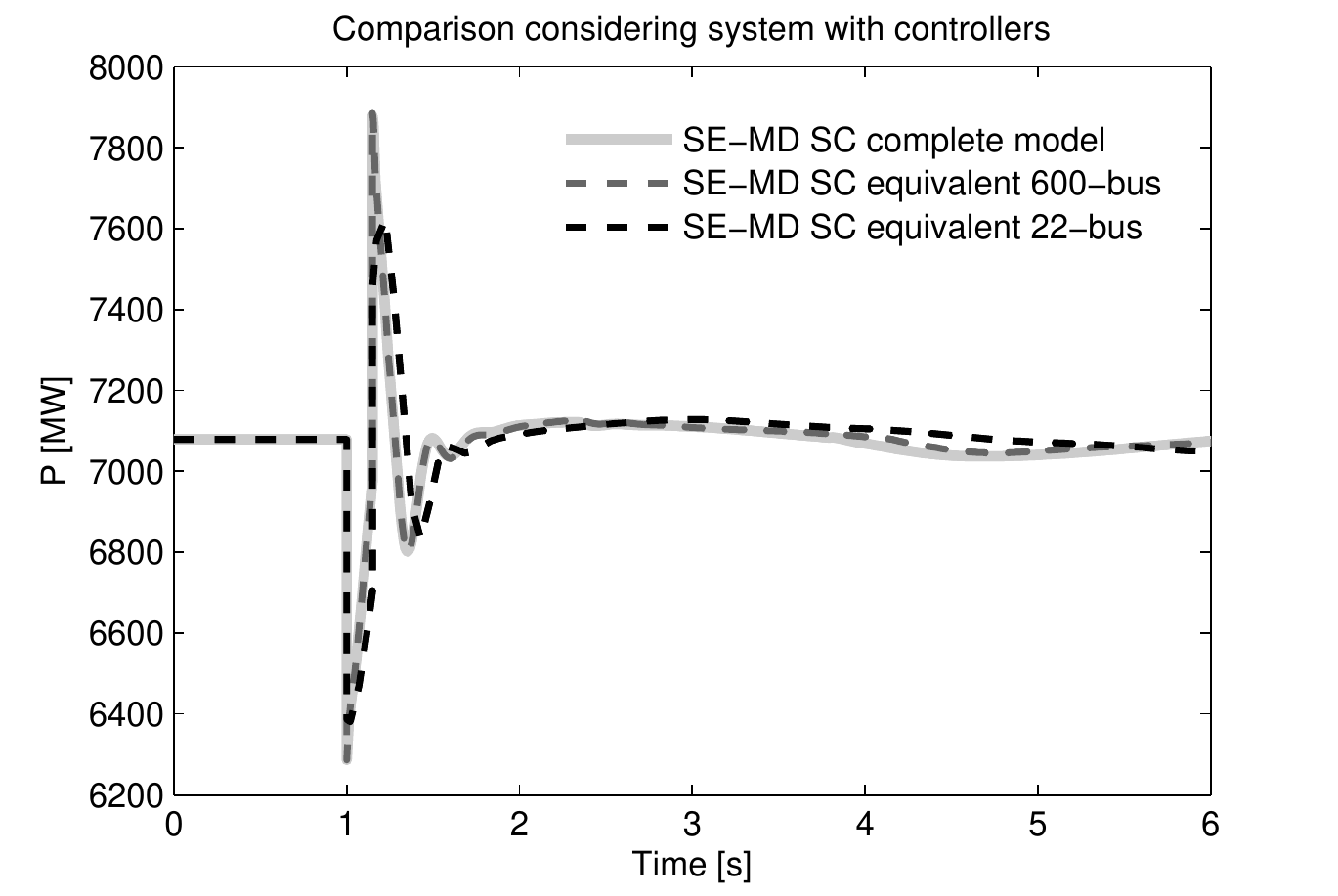}
    \caption{Comparison of the 22-bus and 600-bus equivalent model responses to the one of the complete model.}
     \label{fig:fig12}
\end{figure}

Fig. \ref{fig:fig12} presents a comparison of the responses of the Itaipu generators given by all three models under study. Although some precision is sacrificed in the 22-bus equivalent, the results are still precise enough to reproduce the response of the complete model. The last row of Table \ref{tab:tab01_stat} shows that this 22-bus equivalent enables real-time and accurate simulations of the complete model. 

Several other disturbances were used to calculate the average simulation times for both the 600- and 22-bus equivalent models, but their results are not shown here due to space limitations.  

\begin{figure}[ht]
% \hspace*{-0.6cm}
\centering
    \includegraphics[scale=0.7]{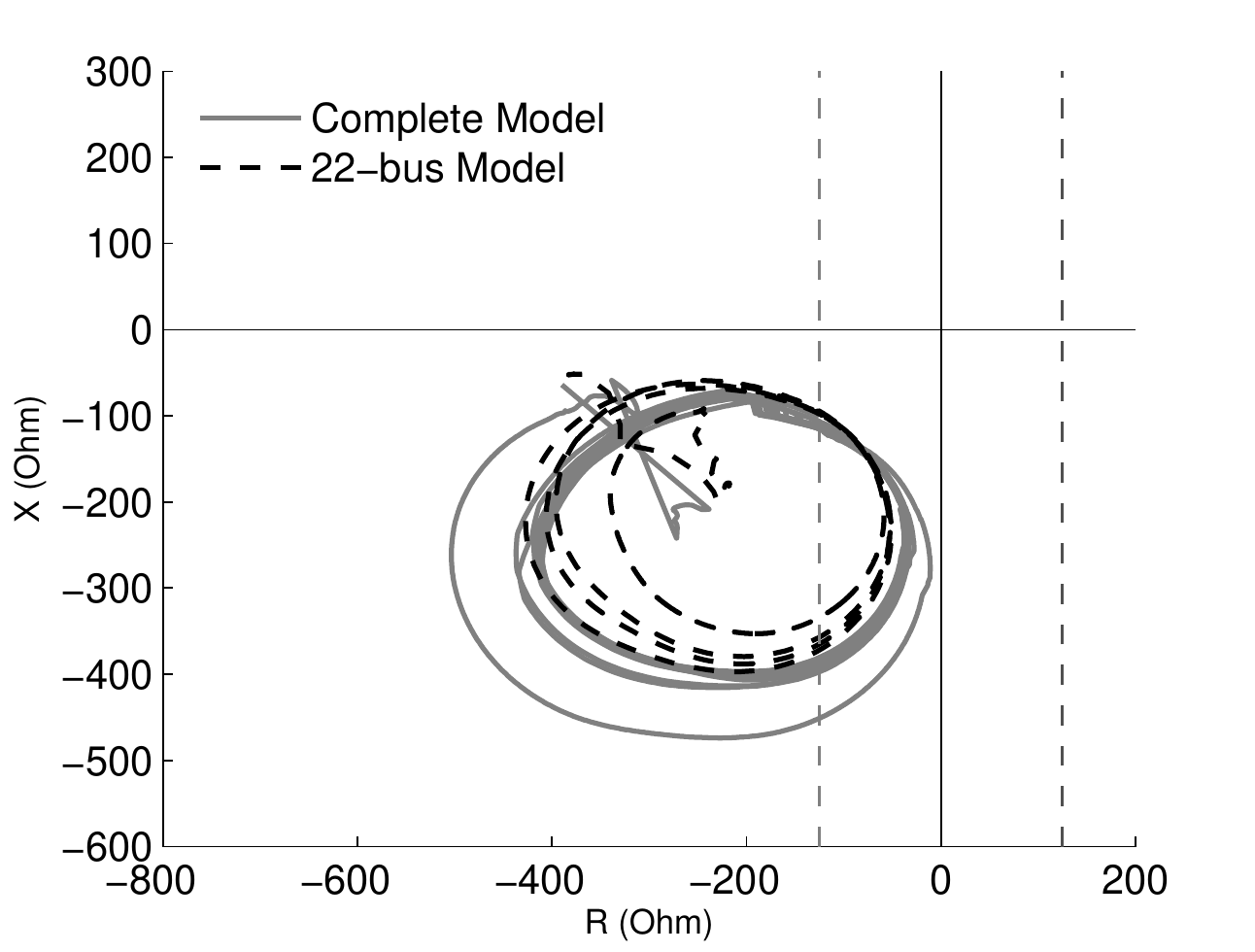}
     \caption{Comparison between the apparent impedance at the AYO substation recorded by the IED in the HIL simulation and by the computer simulation of the complete model.}
     \label{fig:fig13}
\end{figure}

In the real-time simulations, one of the main functions evaluated was the out-of-step protection installed in the SEMD and the AYO substations. Fig. \ref{fig:fig13} presents the comparison of the apparent impedance recorded by the IED in the HIL simulation to the one obtained by the computer simulation of the complete model, once again showing a satisfactory match from the qualitative viewpoint, which enabled batch simulations for protection studies.

\vspace{0.1cm}

\section{Concluding remarks}

This paper proposed a numerical approach to aggregate controllers with input and/or output variables that have distinct physical natures. The application of the proposed approach enabled fast and accurate simulations of large-sized systems while overcoming limitations of other methods related to controller aggregation. This approach was applied to the development of two equivalent models for interconnection studies Paraguayan-Argentinean interconnection studies.

The more accurate representation of aggregated controllers in the equivalent model has proven to be a significant improvement from the accuracy viewpoint. This also resulted in the possibility of finding trade-offs between accuracy and size of the equivalent model, which enabled the creation of reduced order models for hardware-in-the-loop simulations. This feature is extremely important for tuning of protection and control devices to be deployed in large-sized systems. 

Perspectives of future work include the application of this method to aggregate network controllers (such as Static Var Compensators, for example), among others.

%-------------------------------------------------------
\vspace{0.2cm}

\bibliographystyle{IEEEtran}
\bibliography{refs}

\end{document}